\documentclass[journal=amlcef,manuscript=letter]{achemso}

\usepackage{chemformula}
\usepackage[T1]{fontenc}

\author{Hassan Ouhbi}
\author{Ulrich Aschauer}
\email{ulrich.aschauer@dcb.unibe.ch}
\affiliation[University of Bern]
{Department of Chemistry and Biochemistry, University of Bern, Bern, Switzerland}

\title{Nitrogen loss and oxygen evolution reaction activity of perovskite oxynitrides}

\begin{document}

%\begin{tocentry}
%\includegraphics{toc.png}
%\end{tocentry}

\begin{abstract}
  Perovskite oxynitride photocatalysts were reported by experiment to evolve small amounts of N$_2$ due to the self-oxidation of nitrogen ions by photo-generated holes. The N$_2$ evolution rate was observed to decrease with increasing reaction time and was found to be correlated with a decrease in O$_2$ evolution (OER) activity, the origin of this latter effect however being unknown. Here we investigate, by means of density functional theory calculation, anion vacancies at the TaON-terminated (001) surface of the perovskite oxynitride SrTaO$_2$N. We find an energetic preference for oxygen and nitrogen vacancies to reside at the surface, where they are spontaneously healed by *O and *OH adsorbates under OER conditions. For nitrogen vacancies, this self-healing leads to an altered stoichiometry Ta$_4$O$_{8+x}$N$_{4-x}$ that is accompanied by electron doping. Substitution of N by O at the surface also leads to tensile strain, which confines the excess charge to the very surface layer, affecting the binding energy of reaction intermediates and significantly increasing the OER overpotential. This peculiar change in electronic structure thus provides an atomic scale explanation for the experimentally observed drop in OER activity of perovskite oxynitrides.
\end{abstract}

\newpage
Oxynitride perovskites with the general formula ABO$_2$N have emerged as  highly promising water splitting photocatalysts under visible light due to their smaller band gaps compared to pure oxides \cite{Maeda2007, Balaz2013, Grabowska2016}. This reduction in band gap is a result of the partial substitution of oxygen by the less electronegative nitrogen. Tantalate \cite{Maeda2007} as well as other perovskite oxynitrides \cite{Rachel2005} and oxynitrides in other crystal structures \cite{Higashi2011} were reported to evolve N$_2$ during the early stages of the photocatalytic reaction. This is due to the oxidation of the N$^{3-}$ ions by the photo-generated holes as put in evidence by a reduced X-ray photo-electron spectroscopy (XPS) nitrogen signal after the reaction \cite{Minegishi2013}. For LaTiO$_2$N it was shown that the photocurrent significantly dropped within a minute and that the O$_2$ evolution rate gradually decreased with time \cite{Minegishi2013}. While it seems likely that atomic-scale defects formed during N$_2$ evolution are at the origin of the reduced OER activity, there have not been any investigations of the atomic-scale origins for this correlation. 

The role of surface point-defects on the OER activity has been studied theoretically for pure oxides, where for SrCoO$_3$ surfaces oxygen vacancies were reported to reduce the OER activity while they can also be healed by oxygen adsorbates \cite{Tahini2016, Tahini2017}. On the other hand, oxygen vacancies were reported to be stable in the sub-surface of Fe$_2$O$_3$ (001) and to remarkably increase the OER activity \cite{Nguyen2015}. For RuO$_2$ it was shown that defects can either increase or decrease the overpotential, depending on the exact geometry \cite{Dickens2017}. These results show that the defect population can indeed have a marked influence on the OER activity while the exact effect depends on the material and the defect location.

The goal of this work, is to investigate the effect of oxygen and nitrogen vacancies on the OER activity of perovskite oxynitride photocatalysts, using SrTaO$_2$N (001) as a prototypical example. We first study the thermodynamic stability of oxygen and nitrogen vacancies in the three outermost atomic layers, showing a preference of both defects to reside at the surface. Subsequently we show how surface vacancies are healed by oxygen adsorbates under photocatalytic application conditions. For nitrogen vacancies this leads to a change in surface stoichiometry. We find a strong increase of the overpotential with decreasing nitrogen content and rationalise this in terms of an electron enrichment of the surface.

\begin{figure}
	\centering
	\includegraphics{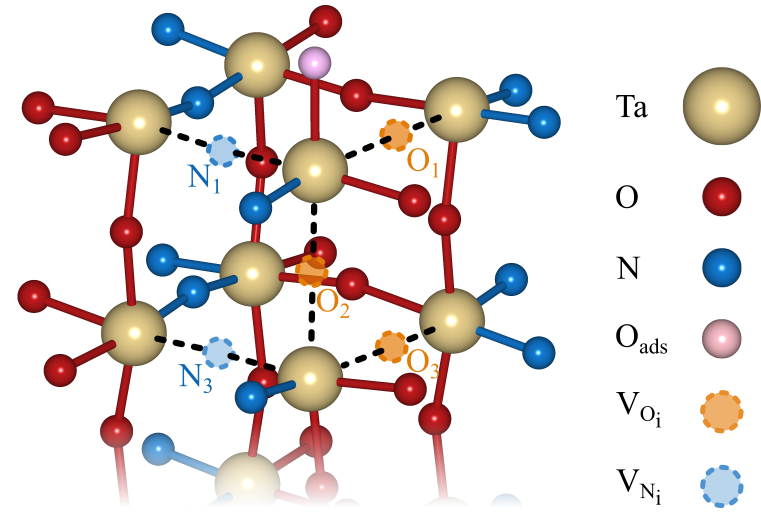}
	\caption{The TaON-terminated (001) surface of SrTaO$_2$N showing the possible anion vacancies in the surface and subsurface layers, where the subscript $i$ indicates the number of the atomic layer.}
	\label{fig:vac_sites}
\end{figure}

We start by investigating the formation energies of O and N vacancies, in the three outermost layer of the TaON terminated SrTaO$_2$N (001) surface (see Figure \ref{fig:vac_sites}). The oxygen and nitrogen vacancies have a charge of +2 and +3 relative to the original lattice site, which are represented as $\mathrm{V^{\bullet\bullet}_O}$ and $\mathrm{V^{\bullet\bullet\bullet}_N}$ in Kr{\"o}ger-Vink notation \cite{Kroger1956}. For simplicity, we will however use $\mathrm{V_O}$ and $\mathrm{V_N}$ in the following. The formation energies $E_f$ of the $\mathrm{V_O}$ and $\mathrm{V_N}$ were computed as 
\begin{equation}
	E_f(\mathrm{V_{O/N}}) = E_\mathrm{tot, V_{O/N}} - E_\mathrm{tot, stoi} + \mu_\mathrm{O/N},
\label{eq:vac_form}
\end{equation}
where $E_\mathrm{tot, stoi}$ and  $E_\mathrm{tot, V_{O/N}}$ are the DFT total energies of the stoichiometric and defective surface respectively and $\mu_\mathrm{O} = \frac{1}{2}E_\mathrm{O_2}$ and $\mu_\text{N} = \frac{1}{2}E_\mathrm{N_2}$ are the chemical potentials of oxygen and nitrogen in equilibrium with $\mathrm{O_2}$ and $\mathrm{N_2}$ gas respectively.

\begin{table}
	\caption{Formation energies of oxygen $\mathrm{V_{O_i}}$ and nitrogen $\mathrm{V_{N_i}}$ vacancies at positions shown in Figure \ref{fig:vac_sites} computed in the oxygen and nitrogen-rich limit respectively.}
	\begin{tabular}{cc}
		\hline
		Vacancy & $\Delta$E$_\text{f}$ (eV)\\
		\hline 
		V$_{\text{O}_\text{1}}$& 2.26\\
		V$_{\text{O}_\text{2}}$& 2.90\\
		V$_{\text{O}_\text{3}}$& 3.27\\
		\hline
		V$_{\text{N}_\text{1}}$& 2.73\\ 
		V$_{\text{N}_\text{3}}$& 3.32\\
		\hline 
	\end{tabular}
	\label{table:vac_formation}
\end{table}

As shown in Table \ref{table:vac_formation}, the formation energies of anion vacancies in the topmost atomic layer $E_f(\mathrm{V_{O_1}})$ and $E_f(\mathrm{V_{N_1}})$ are smaller than the ones in lower layers. Anion vacancies therefore have an energetic preference to reside at the (001) surface compared to subsurface layers. This is likely related to larger structural relaxations, observed for the lower coordinated surface Ta sites adjacent to the defects compared to Ta sites in the 3rd layer. Comparing the absolute values of the $\mathrm{V_O}$ and $\mathrm{V_N}$ formation energies at the chosen chemical potential conditions, we observe an almost equal formation energy in the 3rd layer, while the $\mathrm{V_O}$ is significantly easier to form at the surface compared to a $\mathrm{V_N}$. We note however that under photocatalytic conditions there usually is a high partial pressure of oxygen at the surface due to the OER adsorbates, while the reaction environment is significantly poorer in nitrogen. This may render the formation of $\mathrm{V_N}$ more favourable than the one of $\mathrm{V_O}$.

\begin{figure}
	\centering
	\includegraphics{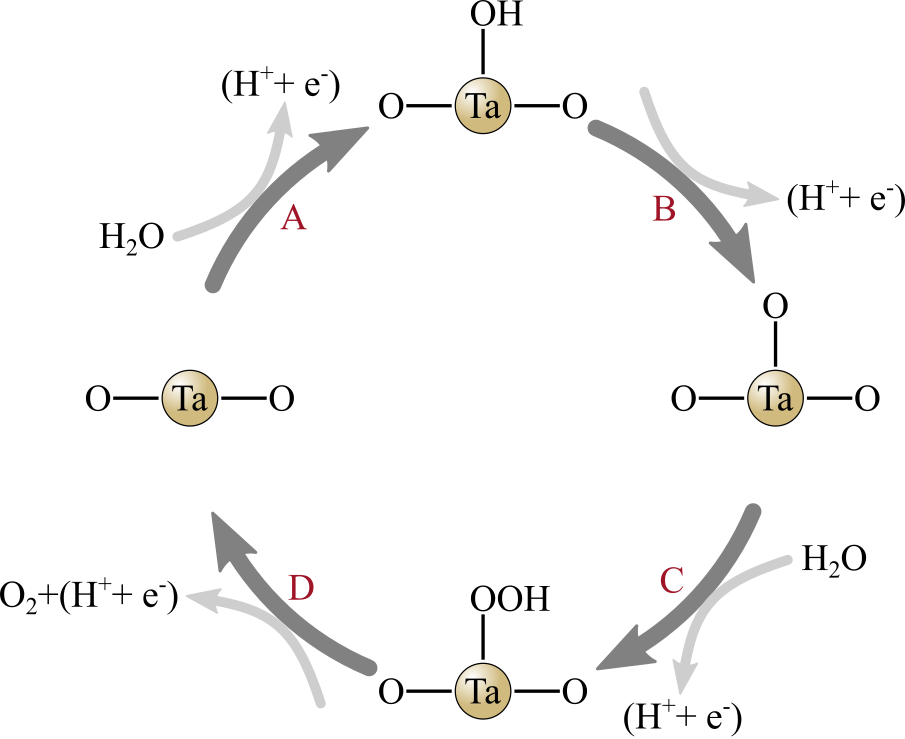}
	\caption{Schematic representation of oxygen evolution reaction mechanism considered in the present work.}
	\label{fig:schematic}
\end{figure}

\begin{figure}
	\centering
	\includegraphics{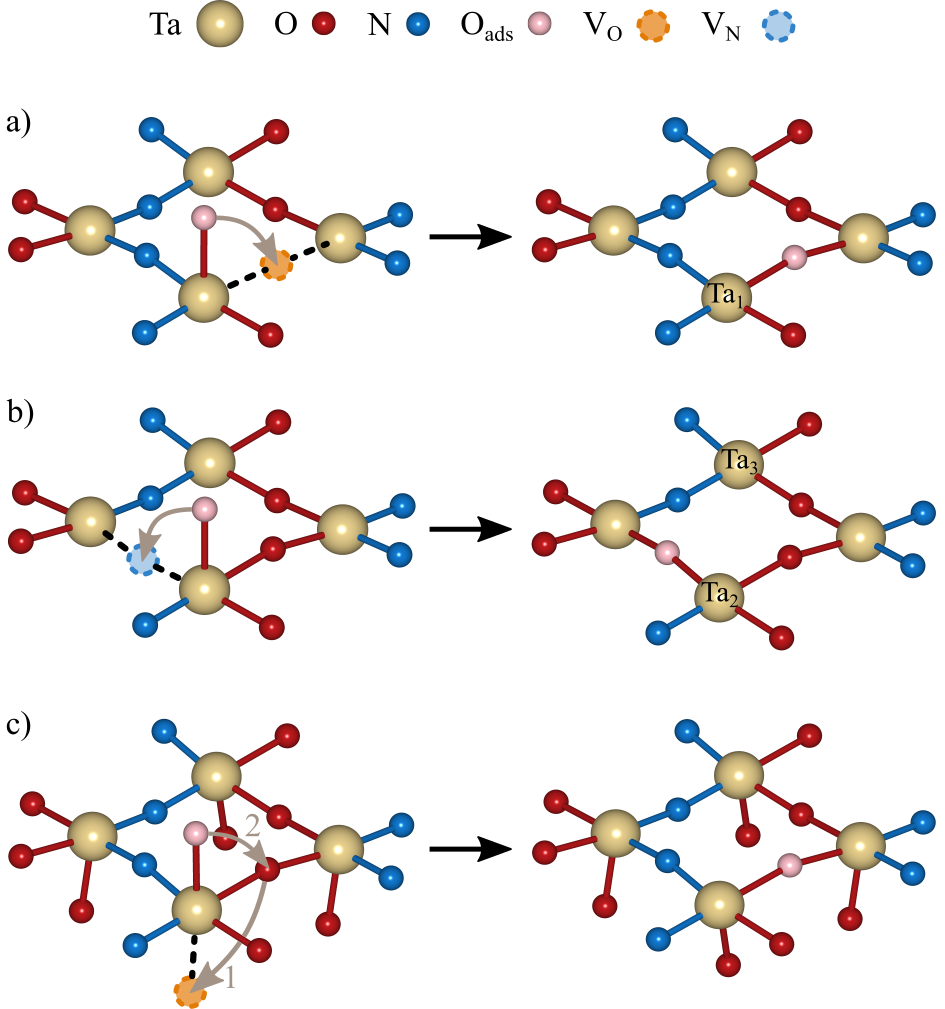}
	\caption{Self-healing of anion vacancies on the TaON terminated (001) surface of SrTaO$_2$N, a) surface oxygen vacancy ($\mathrm{V_{O_1}}$), b) surface nitrogen vacancy ($\mathrm{V_{N_1}}$) and c) sub-surface oxygen vacancy $\mathrm{V_{O_2}}$.}
	\label{fig:self-healing}
\end{figure}

\begin{figure}
	\centering
	\includegraphics{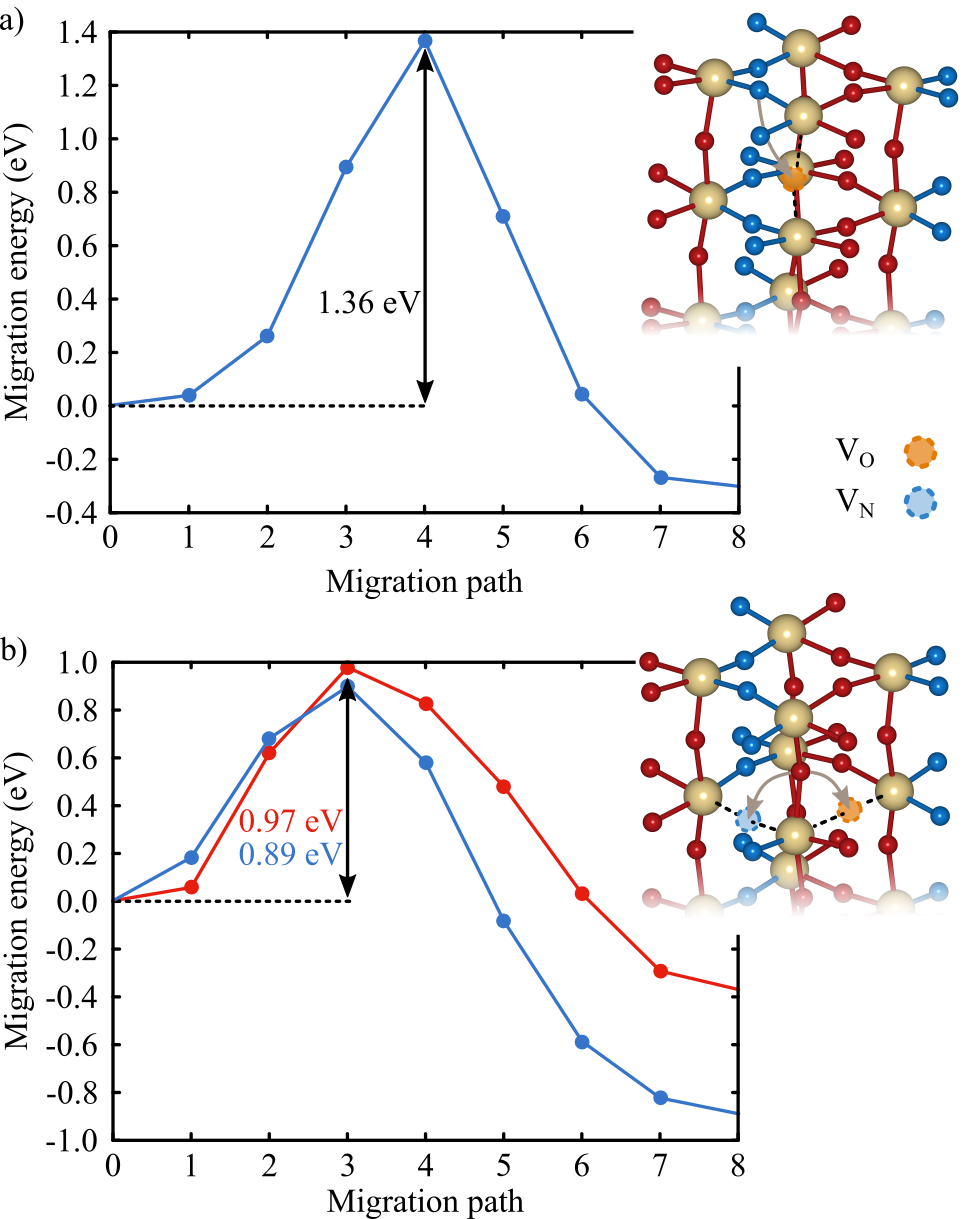}
	\caption{Migration barrier of a) surface nitrogen to a sub-surface oxygen $\mathrm{V_{O_2}}$ and b) sub-surface oxygen to $\mathrm{V_{N_3}}$ (blue) and $\mathrm{V_{O_3}}$ (red).}
	\label{fig:vac_migration}
\end{figure}

 To investigate the effect of these vacancies on the surface chemistry of SrTaO$_2$N, we investigate their interaction with OER adsorbates. Under photo-electrochemical conditions, SrTaO$_2$N (001) surfaces will be covered with *O adsorbates \cite{Ouhbi2018} but during the conventional OER mechanism we consider here (see Figure \ref{fig:schematic}) *OH and *OOH adsorbates will also be present. We observe a spontaneous and barrierless migration of the *O and *OH adsorbates into surface anion vacancies $\mathrm{V_{O_1}}$ and $\mathrm{V_{N_1}}$ (see Figure \ref{fig:self-healing}a and b), while the *OOH adsorbate dissociates into a O and OH fragment, the former migrating into the vacancy and the latter remaining bound to a surface Ta site. For the subsurface $\mathrm{V_{O_2}}$ we observe the spontaneous and barrierless two-step process shown in Figure \ref{fig:self-healing}c), where a surface oxygen migrates into the subsurface vacancy (step 1) followed by migration of an *O adsorbate into the resulting surface vacancy (step 2). While we readily observed this process involving surface oxygen ions for step 1, we did not see it for surface nitrogen ions. As shown by the nudged elastic band (NEB) pathway in Figure \ref{fig:vac_migration}a), this is related to a rather large barrier of 1.36 eV associated with the migration of a nitrogen from the 1st to the 2nd layer. Consequently oxygen vacancies in the 2nd layer can only be healed by surface oxygen ions.

We also performed NEB calculations for the migration of an oxygen in the 2nd layer to a $\mathrm{V_{O_3}}$ or $\mathrm{V_{N_3}}$ in the 3rd layer. As shown in Figure \ref{fig:vac_migration}b), there are barriers of 0.97 eV and 0.89 eV associated with the migration into the $\mathrm{V_{O_3}}$ and $\mathrm{V_{N_3}}$ respectively, which will kinetically hinder this process. Anion vacancies in the 3rd and deeper layers are therefore unlikely to be healed by migration of these defects to the surface.

Since the subsurface anion vacancies $\mathrm{V_{O_3}}$ and $\mathrm{V_{N_3}}$ cannot be healed by surface adsorbates during water oxidation, we investigate their effect on the OER activity. We adopt the OER mechanism schematically shown in Figure \ref{fig:schematic}, where first a water molecule transfers one of its protons to the electrolyte, leading to the formation of an *OH adsorbate (step A). This is followed by step B, in which the *O adsorbate is formed after deprotonation of *OH. In step C, the hydroperoxo group *OOH is formed after dissociation of a second water molecule and transfer of one of its proton to the electrolyte. Finally, the O$_2$ is released from the surface after deprotonation of the *OOH (step D). From the adsorbate binding energies with respect to $\mathrm{H_2O}$ and $\mathrm{H_2}$ and the resulting overpotentials, shown in Table \ref{table:binding}, we see a significantly increase in overpotential from 1.01 V in the stoichiometric case \cite{Ouhbi2018} to 1.81 V and 1.93 V respectively on a surface Ta above the defect for the $\mathrm{V_{N_3}}$ and $\mathrm{V_{O_3}}$. These overpotentials change slightly for Ta sites further from the defect, where we calculate 2.03 V and 1.88 V for the $\mathrm{V_{N_3}}$ and $\mathrm{V_{O_3}}$ respectively. These increased overpotentials compared to the stoichiometric surface are related to a stronger binding of the *OH and *O adsorbates in presence of the vacancies, which manifests as an increased free energy change of step C. The presence of these kinetically stabilised anion vacancies in the 3rd layer thus leads to a significant increase of the overpotential by up to 1 V compared to the defect-free surface.

\begin{table}
	\caption{Binding energy of the OER adsorbates and calculated overpotential at pH = 0 and $U_b$ = 0 in presence of $\mathrm{V_{N_3}}$ and $\mathrm{V_{O_3}}$ defects in the 3rd layer as well as on different Ta$_i$ sites on surface with altered composition due to self-healing shown in Figure \ref{fig:self-healing}.}
	\begin{tabular}{cccccc}
		\hline
		Site & Defect & $\Delta$G$_{OH}$ (eV)& $\Delta$G$_{O}$ (eV) & $\Delta$G$_{OOH}$ (eV) & $\eta$ (V) \\
		\hline
		Ta & V$_{\text{N}_\text{3}}$ & -1.35 & -0.77& 2.27 & 1.81\\
		Ta & V$_{\text{O}_\text{3}}$ & -1.42 & -0.97& 2.19 & 1.93\\
		\hline
		Ta$_1$ & - & -0.45 &  0.94 & 3.19 & 1.02\\
		Ta$_2$ & - & -1.35 & -0.15 & 2.20 & 1.46\\
		Ta$_3$ & - & -1.23 &  0.11 & 2.41 & 1.25\\ 
		\hline 
	\end{tabular}
	\label{table:binding}
\end{table}

In contrast to anion vacancies in the 3rd layer, these in the 1st and 2nd layer ($\mathrm{V_{N_1}}$ and $\mathrm{V_{O_{1,2}}}$) can be healed by OER adsorbates as discussed above. We compute the OER at the different Ta$_i$ sites shown in Figure \ref{fig:self-healing}, to characterise the effect of this change on the activity. From the results reported in Table \ref{table:binding}, we see that the Ta$_1$ site (see Figure \ref{fig:self-healing}a) has an overpotential of 1.02 V, nearly equivalent to the stoichiometric surface \cite{Ouhbi2018}, which is expected as they have the same structure. The same is true when a subsurface $\mathrm{V_{O_2}}$ is healed by the mechanism shown in Figure \ref{fig:self-healing}c). Healing of a surface $\mathrm{V_{N_1}}$ on the other hand leads to an altered local environment with three instead of two surface oxygen ligands for some of the surface Ta sites. As can be seen from the results in Table \ref{table:binding}, the overpotentials at the inequivalent Ta$_2$ and Ta$_3$ sites increases to 1.46 and 1.25 V respectively. Since the coordination environment of Ta$_3$ is not affected but its overpotential changes, the effect of substituting N with O is not local but affects neighbouring transition metal sites as well. We also notice that for both Ta$_2$ and Ta$_3$ step D becomes the overpotential determining step, whereas it was step C on the stoichiometric surface \cite{Ouhbi2018}. Self-healing of $\mathrm{V_{N_1}}$ by OER adsorbates therefore significantly reduces the OER activity, which is due to a stronger bonding of the *O and *OH intermediates at both Ta$_2$ and Ta$_3$ compared to Ta$_1$ on the stoichiometric surface.

\begin{figure}
	\centering
	\includegraphics{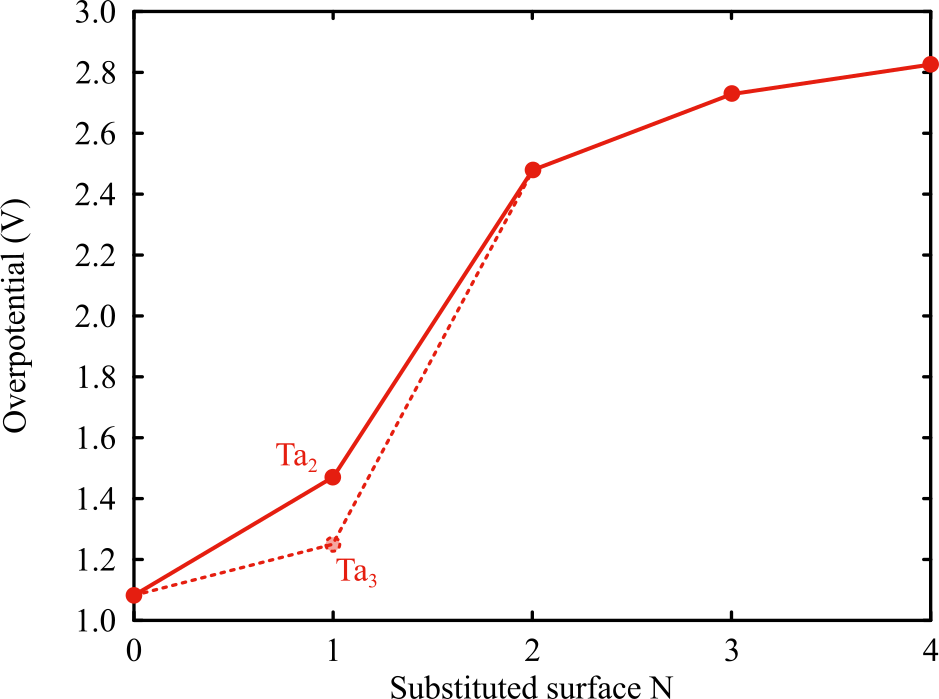}
	\caption{Computed overpotential as a function of the number of substituted surface nitrogen ions. The dashed line indicates the different overpotential at the Ta$_3$ site for the $x$=1 configuration.}
	\label{fig:subs_overpotential}
\end{figure}

The above results show that nitrogen vacancies at the surface are healed by OER adsorbates, changing the stoichiometry of the surface layer to $\mathrm{Ta_4O_{8+x}N_{4-x}}$, where $x$ is the number of substituted surface nitrogen atoms. Given that the OER typically takes place in nitrogen poor environments that promote $\mathrm{V_N}$ formation and since we have shown that the substitution of a single surface nitrogen by oxygen has a marked effect on the predicted overpotential, we compute now the overpotential for different concentrations of substituted surface nitrogen ($x$ = 0, 1, 2, 3, 4). From the results shown in Figure \ref{fig:subs_overpotential}, we observe a continuous increase in overpotential with decreasing number of surface nitrogen ions. It is interesting to note that for $x$=2 and $x$=3 different arrangements of the substitution sites are possible, which however all result in the same overpotential. At these N concentrations, the OER activity does therefore only depend on the overall nitrogen content but not on the exact atomic scale configuration. For $x$=1 we observe a site dependence with different overpotential for Ta$_2$ and Ta$_3$ as discussed above.

\begin{figure}
	\centering
	\includegraphics{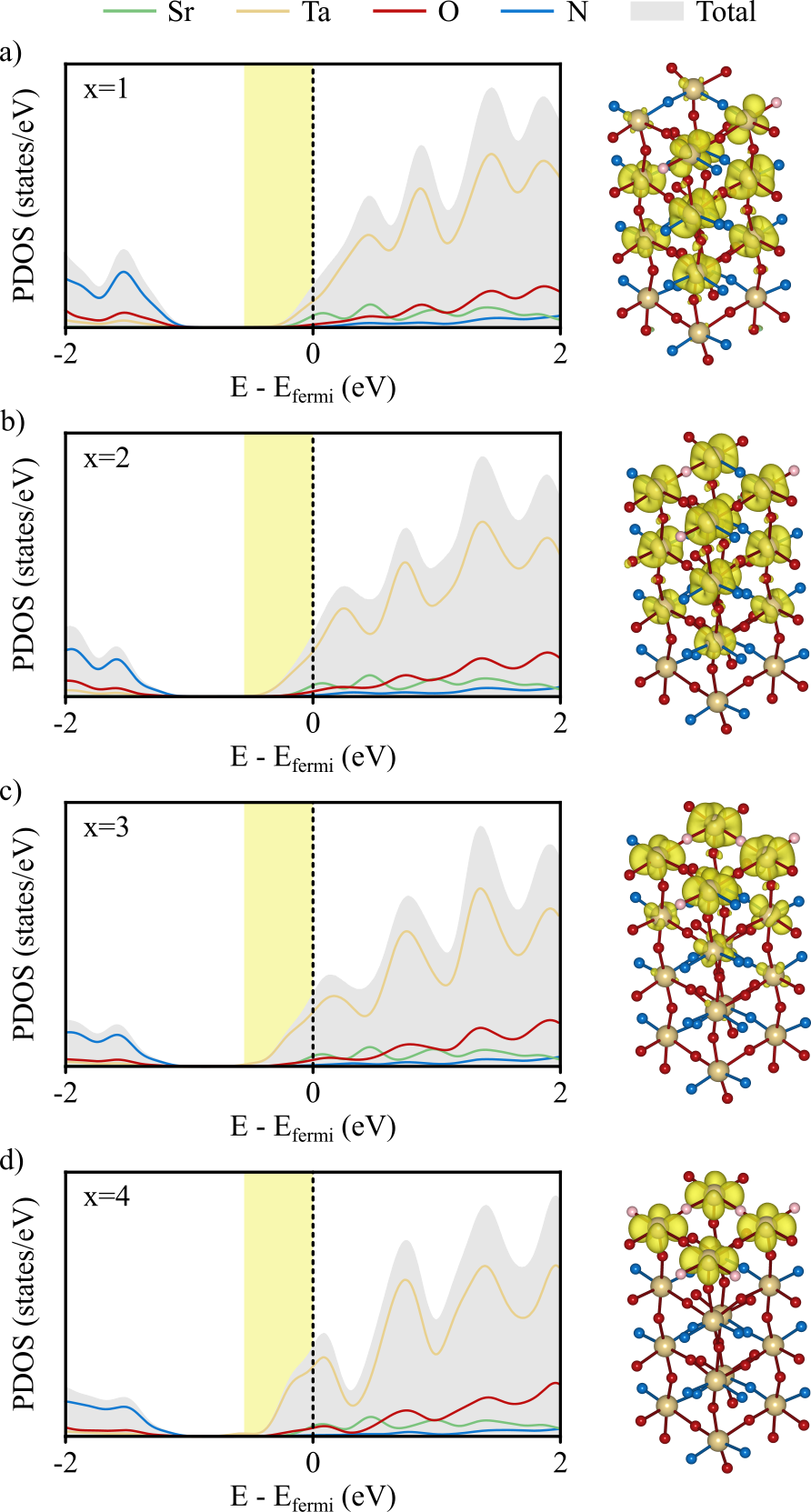}
	\caption{Total and projected density of states for a) 1, b) 2, c) 3 and d) 4 substituted surface nitrogen ions. The isosurfaces show the integrated local density of states in the energy range indicated by the yellow background in the density of states.}
	\label{fig:surface_dos}
\end{figure}

\begin{figure}
	\centering
	\includegraphics{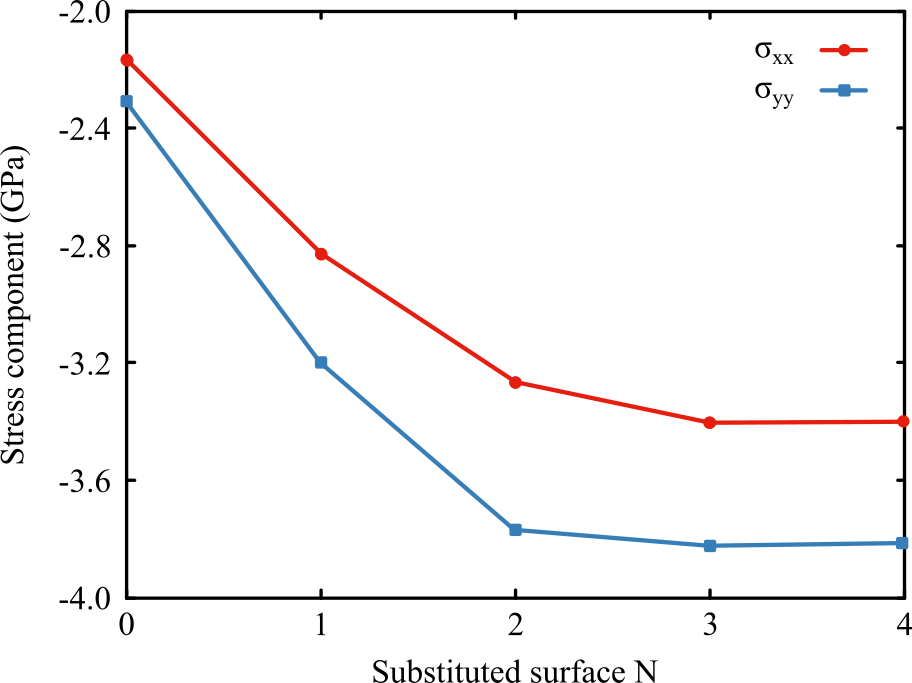}
	\caption{Diagonal components of the stress tensor $\sigma_{xx}$ and $\sigma_{yy}$ as a function the number of substituted surface nitrogen ions.}
	\label{fig:stress}
\end{figure}

From the projected density of states (PDOS) for surfaces with different nitrogen content shown in Figure \ref{fig:surface_dos}, it can be seen that substituting nitrogen by oxygen leads to an accommodation of the excess electrons in conduction band (CB) states, the Fermi energy shifting progressively higher into the Ta-\textit{d} dominated CB as $x$ increases. It is interesting to note that the shape of the DOS at the bottom of the CB changes from band-like at $x$=1 to an increasingly more localised defect-state like feature at higher $x$. This is also reflected by changes in the integrated local density of states (ILDOS) of the occupied CB states below the Fermi energy. For $x$=1 we see that the density is delocalised over the entire slab but with larger contributions on the surface Ta atoms that have an increased number of O ligands, which rationalises the different overpotential for Ta$_2$ and Ta$_3$ sites reported above. For $x$=2 and higher this distinction of surface sites vanishes and in addition we observe an increasing localisation of the excess charge near the surface until it resides only on surface Ta atoms at $x$=4. From the diagonal components $\sigma_{xx}$ and $\sigma_{yy}$ of the stress tensor shown in Figure \ref{fig:stress}, we can see that substitution of N with O leads to a contraction of the surface layer that manifests in compressive stress on the simulation cell. Since this contraction is hindered by the underlying bulk, the surface layer is under tensile strain, where the increased local volume promotes accommodation of the excess charge \cite{Adler2001, Aschauer2013}. The availability of excess charge leads to stronger bonds of the adsorbates, increasing the overpotential and may in addition also lead to increased electron-hole recombination at the surface, which also reduces the efficiency of the photocatalyst.

%\section{Conclusions}

In summary, we have shown a preference for both oxygen and nitrogen vacancies to reside in surface layers, the migration of vacancies from the 3rd layer upwards being associated with large energy barriers of the order of 0.9 eV. Under oxygen evolution reaction (OER) conditions we find that anion vacancies in the first and second layer are spontaneously healed by oxygen adsorbates. Healing of nitrogen vacancies in the surface layer will lead to a change in stoichiometry and an increase of the OER overpotential with decreasing nitrogen content. We can relate this to a tensile-strain induced localisation of excess charge at the surface that leads to a stronger binding of *O and *OH reaction intermediates. Our findings thus show that nitrogen loss from oxynitride surfaces results in a reduced OER activity and provide a rationale for the experimentally observed nitrogen evolution coupled with a progressive decrease in the oxygen evolution for oxynitride photocatalysts. Our results also show that for well-performing oxynitride photocatalysts, it is mandatory to prevent the self-oxidation of nitrogen, for example by co-catalysts, co-doping of donor elements or protective coatings.

%%%%%%%%%%%%%%%%%%%%%%%%%%%%%%%%%%%%%%%%%%%%%%%%%%%%%%%%%%%%%%%%%%%%%%%%%%%%%%%
%\section{Computational Methods}

\newpage
\textbf{Computational Methods}

Our density functional theory (DFT) calculations are performed using the Perdew-Burke-Ernzerhof (PBE) exchange correlation functional \cite{Perdew1996}, as implemented in the Quantum ESPRESSO package \cite{Giannozzi2009}. Ultrasoft pseudopotentials \cite{Vanderbilt1990} with Sr(4s, 4p, 5s), Ta(5s, 5p, 5d, 6s), O(2s, 2p) and N(2s, 2p) valence states were used to describe electron-nuclear interactions and wave functions were expanded in plane waves up to a kinetic energy cutoff of 40 Ry combined with 320 Ry for the augmented density.

We construct the TaON-terminated (001) slab (lateral dimensions 8.182 $\times$ 8.182 \AA) with 8 TaON/SrO layers and an in-plane \textit{cis} order of the nitrogen ions based on the most stable SrTaO$_2$N bulk structure \cite{Ouhbi2018}. Periodic images of the slab along the surface normal direction are separated by a vacuum of 15 \AA, the bottom two atomic layers were fixed at bulk positions and a dipole correction \cite{Bengtsson1999} was introduced along the z-direction to cancel spurious electric fields across the vacuum. Reciprocal space integration is performed with a 4$\times$4$\times$1 k-point mesh \cite{Pack1977} and structures were relaxed with force and total energy thresholds of 10$^{-3}$ eV\AA$^{-1}$ and 10$^{-6}$ eV respectively.

The thermochemical scheme developed by N\o rskov and co-workers \cite{Norskov2004, Valdes2008} is used to calculate the free energies of the OER proton-coupled electron transfer (PCET) steps under applied potential and pH conditions. The free energy change of a reaction step $\Delta G(U_b, \mathrm{pH})$ is calculated as
\begin{equation}
\Delta G(U_b, \mathrm{pH}) = \Delta E + (\Delta\mathrm{ZPE} - T\Delta S) - \mathrm{e}U_b - k_BT\cdot \ln(10)\cdot \mathrm{pH}
\label{eq:deltaG}
\end{equation}
where $\Delta E$ is the reaction energy calculated by DFT, $-\mathrm{e}U$ is the energy shift due to the applied potential, $-k_BT\cdot \ln(10)\cdot \mathrm{pH}$ is the free-energy shift at a given pH and $\Delta\mathrm{ZPE}$ and $\Delta S$ are the changes in zero point energy and entropy of the reaction intermediates respectively, which were taken from our previous work \cite{Ouhbi2018}. The thermodynamic overpotential $\eta$, which we use to characterize the OER activity of a given surface, is defined as the potential for which all PCET steps have $\Delta G(U_b, \mathrm{pH})$ smaller than zero relative to the equilibrium potential of 1.23 V:
\begin{equation}
	\eta = \frac{\max(\Delta G)}{\mathrm{e}} - 1.23 \mathrm{V}
\label{eq:overpotential}
\end{equation}
Activation barriers for thermally activated processes are calculated using the climbing image nudged elastic band (CI-NEB) method \cite{Henkelman2000}. 

\begin{acknowledgement}

This work was funded by SNF Professorship Grant PP00P2\_157615. Calculations were performed on UBELIX (http://www.id.unibe.ch/hpc), the HPC cluster at the University of Bern.

\end{acknowledgement}

%\begin{suppinfo}
%
%A listing of the contents of each file supplied as Supporting Information
%should be included. For instructions on what should be included in the
%Supporting Information as well as how to prepare this material for
%publications, refer to the journal's Instructions for Authors.
%
%The following files are available free of charge.
%\begin{itemize}
%  \item Filename: brief description
%  \item Filename: brief description
%\end{itemize}
%
%\end{suppinfo}

%%%%%%%%%%%%%%%%%%%%%%%%%%%%%%%%%%%%%%%%%%%%%%%%%%%%%%%%%%%%%%%%%%%%%
%% The appropriate \bibliography command should be placed here.
%% Notice that the class file automatically sets \bibliographystyle
%% and also names the section correctly.
%%%%%%%%%%%%%%%%%%%%%%%%%%%%%%%%%%%%%%%%%%%%%%%%%%%%%%%%%%%%%%%%%%%%%
\bibliography{library}

\end{document}